\documentclass[prd,aps,preprintnumbers,amsmath,amssymb,nofootinbib,eqsecnum, preprintnumbers]{revtex4-2}
 
\usepackage{graphicx}
 
\usepackage{amsmath}
\usepackage{epstopdf}
 
\usepackage{color}

\usepackage{footnote}
\usepackage{graphicx}
\usepackage{hyperref}
\usepackage{amssymb,epsf}
 
\usepackage{epsfig}
\begin{document}

\title{Non-perturbative correction on the black hole geometry}

\author{Behnam Pourhassan$^1$}\email{b.pourhassan@du.ac.ir}
\author{Hoda Farahani$^{1,2}$} 
\email{h.farahani@umz.ac.ir}
\author{Farideh Kazemian$^{1}$} 
\author{\.{I}zzet Sakall{\i}$^{3}$}
\email{izzet.sakalli@emu.edu.tr}
\author{Sudhaker Upadhyay$^{4, 1}$\footnote{Corresponding author}\footnote{Visiting Associate, IUCAA, Pune, Maharashtra 411007, India}}
\email{sudhakerupadhyay@gmail.com}
\author{Dharm Veer Singh$^5$\footnote{Visiting Associate, IUCAA, Pune, Maharashtra 411007, India}}
\email{veerdsingh@gmail.com}

\affiliation{$^1$School of Physics, Damghan University, Damghan 3671645667, Iran.}
\affiliation{$^2$Center for Theoretical Physics, Khazar University, 41 Mehseti Street, Baku AZ1096, Azerbaijan}
\affiliation{$^3$Physics Department, Eastern Mediterranean University, Famagusta 99628, North Cyprus via Mersin 10, Turkey.}

\affiliation{$^4$Department of Physics, K. L. S. College, Magadh University, Nawada, Bihar 805110, India.}

\affiliation{$^5$Department of Physics, Institute of Applied Sciences and Humanities, GLA University, Mathura 281406, Uttar Pradesh, India.}

\begin{abstract}
In this paper, we use the holographic principle to obtain a modified metric of black holes that reproduces the exponentially corrected entropy. The exponential correction of the black hole entropy comes from non-perturbative corrections. It interprets as a quantum effect which affects black hole thermodynamics especially in the infinitesimal scales. Hence, it may affect black hole stability at the final stage.
Then, we study modified thermodynamics due to the non-perturbative corrections and calculate thermodynamics quantities of several non-rotating black holes.
\end{abstract}

%



\maketitle

\tableofcontents

\section{Introduction}
\label{intro}
A black hole entropy is proportional to the event horizon area \cite{H1,H2}. Thanks to the holographic principle, we can obtain black hole thermodynamics by using its horizon information. It is one of the best ways to study black hole physics. It gives us some important information about black hole evaporation, black hole stability, critical points, and phase transition of black holes. So, in the last decades, there have been several works where various kinds of black hole thermodynamics have been studied.\\
As a black hole size is reduced due to the Hawking radiation, the thermal fluctuations become important and black hole entropy may corrected by a logarithmic term at leading order \cite{2,SPR,Sud}. This logarithmic correction (perturbative correction) on the black hole entropy will modify the black hole thermodynamics so there are several works where mentioned perturbative correction have been investigated \cite{0001,0002,0003,0004,0005,p1,p2,p3, jhap}, which is also extended to other black objects like black rings \cite{PMR}. These corrections may be used to test the quantum theory of gravity \cite{dumb}. However, by reducing more in the black hole size, the non-perturbative corrections become dominant \cite{main}. An exponential term gives the non-perturbative correction on the black hole entropy,
$$S=S_{0}+e^{-S_{0}},$$
where $S_{0}$ is the uncorrected black hole entropy proportional to the horizon radius. It is a universal behaviour that can be used to study various black hole thermodynamics. Recently, modified thermodynamics of some black holes due to the exponential correction has been investigated. For example, some non-rotating black hole thermodynamics were studied by Ref. \cite{N0}, and Myers-Perry black holes were considered by Ref. \cite{N11}. The exponential correction term is negligible at a large radius, while it is dominant at infinitesimal radii. Such correction already considered in literature \cite{n1,n2,n3,n4,n5}. Also in Ref. \cite{Chin} non-perturbative quantum correction is considered to study thermodynamics of the Hořava–Lifshitz black hole. Above corrected entropy assumed in Ref. \cite{Chin}, while the black hole temperature remains unaffected by corrections. Therefore, the study of quantum corrections to the thermodynamics of black branes has gained significant attention in recent research. In Ref. \cite{N22}, both perturbative and non-perturbative corrections to the thermodynamics of black branes have been investigated and observed that the mentioned corrections will modify the relation between the entropy and area of black branes. The use of holographic relations has been employed to obtain the corrected black hole temperature directly from the black hole metrics. Furthermore, the effects of these corrections on various thermodynamic properties have been discussed. The holographic principle, particularly in the context of AdS/CFT and the holographic interpretation of black holes, has been a significant area of study in understanding the thermodynamic properties of black objects \cite{Zaf}. The findings from these studies provide valuable insights into the impact of quantum corrections on the thermodynamics of black branes and the associated implications for black hole entropy and temperature.

This paper is dedicated to identifying an appropriate black hole metric capable of yielding the aforementioned exponential corrected entropy. Employing holographic relations, we aim to derive the corrected black hole temperature directly from the black hole metrics. To achieve this, the original approach of this paper draws upon the methodology outlined in Ref. \cite{1310.4319} to propose modified metrics for certain well-known black holes. These modifications in the metrics are intended to generate the exponential corrected entropy corresponding to the respective black holes. The holographic method employed in this study commences with the formulation of the partition function within the canonical ensemble. Subsequently, we utilize this partition function to compute the thermal average of an operator. This involves considering imaginary time evolution and implementing periodic boundary conditions in the Hilbert space. The outcome of this process provides us with the Hawking temperature of a black hole. Utilizing this obtained temperature in the first law of thermodynamics allows us to extract the black hole entropy. The resultant entropy is anticipated to align with the area entropy relation and is intrinsically linked to the entropy derived from the central charge in quantum field theory. Consequently, by employing the modified metric for black holes, we can establish all the modified thermodynamic relations, thus encompassing the critical aspects of quantum gravity within this framework. Recent developments in holography and black hole thermodynamics have provided valuable insights into the quantum nature of black holes and the interplay of entropy, area, and the holographic principle. The holographic principle, which states that the entropy of ordinary mass is proportional to surface area, has been a fundamental concept in understanding black hole entropy and its relation to quantum gravity. Additionally, studies have focused on non-perturbative quantum corrections and their implications for black hole thermodynamics, including the computation of black hole entropy using the Bekenstein-Hawking formula \cite{karar}.\\
So, in the next section, we present our general method to obtain the modified metric function for the spherical symmetric black holes. Then we examine our method for the $D$-dimensional Schwarzschild black holes in section \ref{G}, Reissner-Nordstr\"{o}m black hole in section \ref{RNBH}, $4D$ Schwarzschild-AdS black hole in section \ref{SABH}, a Charged AdS black hole in section \ref{CABH}, and STU black hole in section \ref{STU}. Finally, we give a conclusion and summary of results in section \ref{C}.

\section{General Formalism}\label{G}
In this section, we would like to introduce our method to produce the exponentially corrected entropy of black holes. Here, we consider a general black hole metric with the Minkowski geometry, which is given by,
\begin{equation}\label{metric-general}
ds^{2}=-g(r)\left(f(r)dt^{2}+dX^{2}\right)+\frac{dr^{2}}{h(r)},
\end{equation}
where $g(r)$, $f(r)$ and $h(r)$ are the metric functions. For the $D$ dimensional space-time, $dX^{2}$ is a $D-2$ dimensional metric, which may be a flat, open or closed space-time. Therefore, the black hole temperature is given by \cite{1310.4319},
\begin{equation}\label{temperature-general}
\frac{1}{T}=\frac{4\pi}{\sqrt{g(r_{0})f^{\prime}(r_{0})h^{\prime}(r_{0})}},
\end{equation}
where $r_{0}$ is the event horizon radius. We would like to use the first law of black hole thermodynamics to obtain the black hole entropy. The first law of black hole thermodynamics in general form is given by \cite{mann},
\begin{equation}\label{First law-general}
dM=TdS+\Omega dJ+\Phi dQ+VdP,
\end{equation}
where $M$, $T$, $S$, $J$, $Q$, $V$ and $P$ are the black hole mass, temperature, entropy, angular momentum, electric charge, volume and pressure, respectively. Also, the rotational parameter $\Omega$ and potential $\Phi$ are conjugate variables of angular momentum and electric charge, respectively. Hence, one can write the following thermodynamics relations:
\begin{equation}\label{th1}
\frac{1}{T}=\left(\frac{dS}{dM}\right)_{J,Q,P},
\end{equation}
\begin{equation}\label{th2}
\frac{1}{\Omega}=\left(\frac{dJ}{dM}\right)_{S,Q,P},
\end{equation}
\begin{equation}\label{th3}
\frac{1}{\Phi}=\left(\frac{dQ}{dM}\right)_{S,J,P},
\end{equation}
\begin{equation}\label{th4}
\frac{1}{V}=\left(\frac{dP}{dM}\right)_{S,J,Q}.
\end{equation}
Combining the equations (\ref{temperature-general}) and (\ref{th1}) gives,
\begin{equation}\label{temperature-th-general}
\left(\frac{dS}{dM}\right)_{J,Q,P}=\frac{4\pi}{\sqrt{g(r_{0})f^{\prime}(r_{0})h^{\prime}(r_{0})}}.
\end{equation}
Hence, the black hole entropy is related to the metric functions. In most spherically symmetric black holes, we have $g(r)=1$, and $f(r)=h(r)$. In that case, using the equation (\ref{temperature-th-general}) one can write,
\begin{equation}\label{temperature-th-general-special}
f^{\prime}(r_{0})=\frac{4\pi}{\left(\frac{dS}{dM}\right)_{J,Q,P}}.
\end{equation}
This paper aims to find the appropriate metric function $f(r)$, reproducing the exponential corrected entropy. We do that for some kinds of black holes. As Ref. \cite{1310.4319} mentioned, the obtained black hole entropy using this method is exactly equal to the area entropy formula.

On the other hand, a general spherically symmetric metric is given by
\begin{equation}\label{metric}
ds^{2}=-f(r)dt^{2}+\frac{dr^{2}}{f(r)}+r^{2}d\Omega_{D-2}^{2},
\end{equation}
where
$d\Omega_{D-2}^{2}$ is the metric of ${D-2}$ dimensional unit sphere with area
\begin{equation}\label{area1}
\Omega_{D-2}=\frac{2{\pi}^{\frac{D-1}{2}}}{\Gamma\left({\frac{D-1}{2}}\right)}.
\end{equation}
According (\ref{metric}), the outer horizon is located at $ f(r=r_{0})=0$.
It yields to the following uncorrected entropy,
\begin{equation}\label{entropy}
S_{0}=\frac{\Omega_{D-2}r_{0}^{D-2}}{4G_{D}}.
\end{equation}
where the indices $0$ denote uncorrected temperature. Also, there is a relation for Hawking temperature based on the first law of thermodynamics that means:
\begin{equation}\label{temperature1}
\frac{1}{T_{H}}=\frac{dS(M)}{dM},
\end{equation}
where $S(M)$ and $T_{H}$ are the black hole entropy and Hawking temperature modified due to the non-perturbative correction. Recently, it was argued that non-perturbative corrections yield the exponential contribution to the black hole entropy \cite{main}. It is indeed dominant at small radii. Using the above relations, we would like to find a modified metric function which generates the following corrected entropy,
\begin{equation}\label{corrected}
S=S_{0}+\eta e^{-S_{0}},
\end{equation}
where $\eta$ is the correction coefficient.\\
The specific heat sign analysis is one of the main tools to obtain information about the black hole's stability.
The specific heat may be given by:
\begin{equation}
C=\frac{-[{S}^{\prime}(M)]^{2}}{{{S}^{\prime\prime}}(M)}
\end{equation}
where prime is derivative concerning $M$. Moreover, by using the Helmholtz free energy and the black hole volume,
\begin{equation}\label{volume}
V=\frac{\pi^{\frac{D}{2}}}{(\frac{D}{2})!} r_{0}^{D},
\end{equation}
one can obtain the thermodynamics pressure.\\
Having pressure helps us to investigate the critical points using the two below conditions,
\begin{equation}\label{condition1}
\left( \frac{\partial P}{\partial V}\right) _{T}=0,
\end{equation}
and
\begin{equation}\label{condition2}
\left( \frac{\partial^{2} P}{\partial V^{2}}\right) _{T}=0.
\end{equation}
We can also calculate the isothermal compressibility
\begin{equation}
\kappa_{T}=-\frac{1}{V}\left(\frac{\partial V}{\partial P}\right)_{T}.
\end{equation}
The change of $E$ in an isothermal expansion is
\begin{equation}\label{iso-S}
\triangle E=\int_{V_{1}}^{V_{2}}{\left[T\left(\frac{\partial P}{\partial T}\right)_{V}-P\right]dV}.
\end{equation}
At zero specific heat, the internal energy remains constant:
\begin{equation}\label{C-E}
C=\frac{\partial E}{\partial T}.
\end{equation}
Hence $\triangle E=0$. In that case, the Joule coefficient
\begin{equation}\label{J-S}
\mu_{J}=\left(\frac{\partial T}{\partial V}\right)_{E},
\end{equation}
is useful to find the situation of black hole temperature in a Joule expansion. If $\mu_{J}=0$, then the temperature is unchanged in a Joule expansion. On the other hand, the negative (positive) sign of the Joule coefficient shows the cooling (warming) of the black hole.

In the case of cooling or warming, one can define the Joule-Kelvin coefficient \cite{J1},
\begin{equation}\label{J-S22}
\mu_{JK}=\left(\frac{\partial T}{\partial P}\right)_{H},
\end{equation}
where $H$ is the black hole enthalpy given by
\begin{equation}\label{H-S}
H=E+PV.
\end{equation}
Finally, it is valuable to mention that the entropy obtained in this method is related to the entropy in the dual field theory \cite{1310.4319}. Hence, we may obtain quantum corrected entropy of the CFT side in the AdS/CFT correspondence by using the modified entropy.
Now, let us try the above general formalism for famous black holes.

\section{$4D$ Reissner-Nordstr\"{o}m Black Hole}\label{RNBH}
Reissner-Nordstr\"{o}m (RN) black hole in four-dimensional space-time is a charged black hole described by the following metric \cite{mann},
\begin{equation}\label{metric-RN}
ds^{2}=-f(r)dt^{2}+\frac{dr^{2}}{f(r)}+r^{2}d\Omega_{2}^{2},
\end{equation}
where $d\Omega_{2}^{2}=d\theta^{2}+\sin^{2}\theta d\varphi^{2}$, and the metric function is given by,
\begin{equation}\label{metric2}
f(r)=1-\frac{2M}{r}+\frac{Q^{2}}{r^{2}},
\end{equation}
where $Q$ is the electric charge.

The mass, Hawking temperature and entropy of RN black hole are:
\begin{equation}\label{temperature2}
M=\frac{r_{0}}{2}(1+\frac{Q^{2}}{r_{0}^{2}}),\ \ \ \ \ \ \ T_{H}=\frac{r_{0}^{2}-Q^{2}}{4\pi r_{0}^{3}},\ \ \ \ \ \ \ S_{0}=\pi r_{0}^{2},
\end{equation}
where $r_{0}=M+\sqrt{M^{2}-Q^{2}}$ is the event horizon radius.
If we assume
\begin{equation}\label{metric modify2}
f(r)=1-\frac{2M^{\prime}}{r}+\frac{Q^{2}}{r^{2}},
\end{equation}
and put $ M^{\prime}\longrightarrow(1-\alpha)M $, where $\alpha=\eta e^{-S_{0}} $ then with the equation (\ref{temperature-th-general-special}) we obtain
\begin{equation}\label{Temperature modify2}
 T_{H}=\frac{r_{0}^{2}-Q^{2}}{(1-\alpha)4\pi r_{0}^{3}}
\end{equation}
and using $ S=\int T_{H}^{-1}dM  $ ($P,Q,J$ are constant) we reproduce the following corrected entropy of RN black hole as follows,
\begin{equation}\label{entropy modify2}
 S=\int T_{H}^{-1}\left(\frac{\partial M^{\prime}}{\partial r_{0}}\right)dr_{0}=S_{0}+\eta e^{-S_{0}}.
\end{equation}
Now, we can study the consequences of this exponential correction.
\subsection{Thermodynamic quantities}
\begin{figure}
\resizebox{0.40\textwidth}{!}
{%
\includegraphics{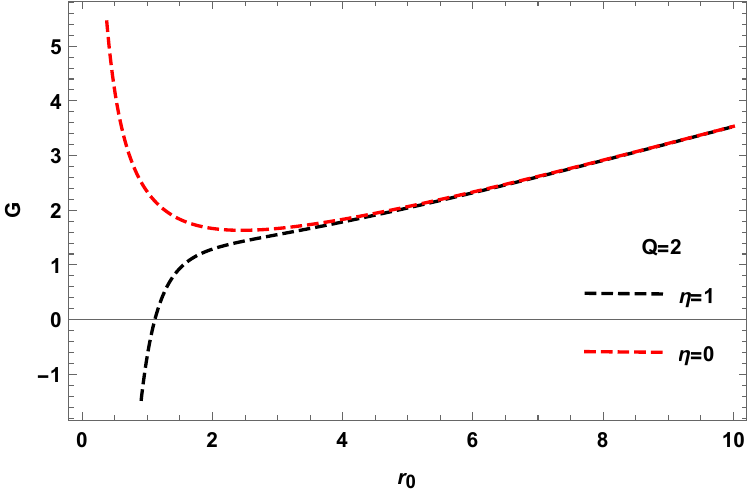}}\\
\resizebox{0.40\textwidth}{!}
{%
\includegraphics{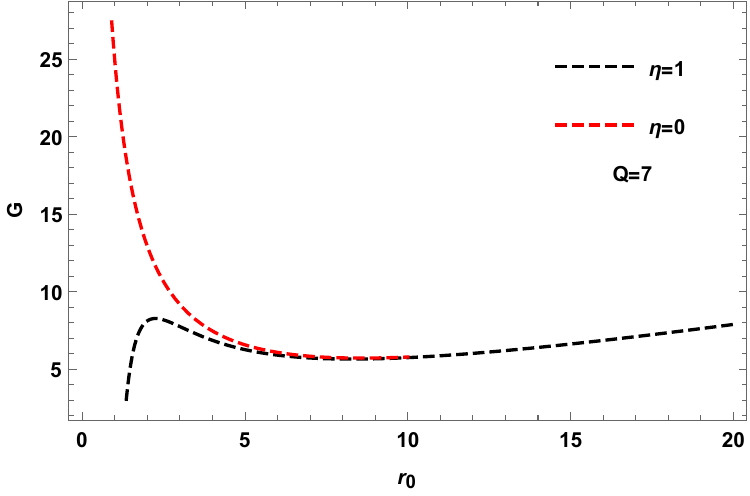}}
\vspace{0.5cm}       
\caption{Gibbs free energy of RN black hole versus $r_{0}$ for $Q=2$, $Q=7$.}
\label{fig:GRN}       
\end{figure}

Helmholtz free energy is calculated as follows,
\begin{eqnarray}\label{Helm2}
F&=&-\int S dT_{H}=\frac{r_{0}^{2}+3Q^{2}}{4r_{0}}\nonumber\\
&+&\frac{\eta}{4\pi r_{0}^{3}}( e^{-\pi r_{0}^{2}}(-r_{0}^{2}+Q^{2}(1+\pi r_{0}^{2})).
\end{eqnarray}
Now, we can easily find the equation of state,
\begin{eqnarray}\label{EOSr2}
&&P=\left( \frac{\partial F}{\partial V}\right)_{T}=
\frac{r_{0}^{2}-3Q^{2}}{16\pi r_{0}^{4}}\nonumber\\
&&+\eta e^{-\pi r_{0}^{2}}\left( \frac{r_{0}^{2}-3Q^{2}}{16\pi r_{0}^{4}}\right) \left( \frac{1}{\pi r_{0}^{2}}(1+\pi r_{0}^{2})\right).
\end{eqnarray}
Then, the Gibbs free energy is calculated as,
\begin{eqnarray}\label{Gibs2}
G&=&\frac{2r_{0}^{2}+3Q^{2}}{6r_{0}}\nonumber\\
&+&\frac{\eta}{4\pi r_{0}^{3}}
(e^{-\pi r_{0}^{2}}(-r_{0}^{2}+Q^{2}(1+\pi r_{0}^{2}))\nonumber\\
&+&\eta e^{-\pi r_{0}^{2}}\left( \frac{r_{0}^{2}-3Q^{2}}{12 r_{0}}\right) \left( \frac{1+\pi r_{0}^{2}}{\pi r_{0}^{2}}\right)
\end{eqnarray}
Neglecting the exponential correction at $Q\rightarrow0$ limit one can find $G=\frac{M+\sqrt{M^{2}-Q^{2}}}{3}$. In plots of Fig. \ref{fig:GRN} we draw Gibbs free energy for some values of the black hole charge to see the effect of the exponential correction.    {  The Gibbs free energy demonstrates a reduction owing to non-perturbative corrections, resulting in a negative value for infinitesimal radii. Conversely, the curves converge for larger radii, indicating that non-perturbative corrections exhibit minimal significance for sizable black holes.}

\subsection{Critical points}
If we put $ r\longrightarrow\left( \frac{3V}{4\pi}\right) ^{\frac{1}{3}} $ and rescaling $ V^{\frac{1}{3}}\longrightarrow V $ in the equation (\ref{EOSr2}) then we can find the following equation of state:
\begin{eqnarray}\label{EOS2}
P&=&\frac{a_{1}Q^{2}+a_{2}V^{2}}{6V^{4}}+\nonumber\\
&&\eta\frac{a_{1}Q^{2}+a_{2}V^{2}}{6V^{4}}\left( \frac{e^{a_{3}V^{2}}(a_{4}+a_{5}V^{2})}{3^{\frac{2}{3}}V^{2}}\right)
\end{eqnarray}
where
\begin{eqnarray}\label{EOS2-2}
a_{1}&=&-3(\frac{\pi}{6})^{\frac{1}{3}}\ \ \ \ \ a_{2}=\frac{(\frac{3}{2\pi})^{\frac{1}{3}}}{2\times 2^{\frac{1}{3}}}\ \ \ \ \ a_{3}=\frac{-3^{\frac{2}{3}}}{2}(\frac{\pi}{2})^{\frac{1}{3}}\nonumber\\
a_{4}&=&2(\frac{2}{\pi})^{\frac{1}{3}}\ \ \ \ \ \ \ \
a_{5}=3^{\frac{2}{3}}.
\end{eqnarray}
Applying the conditions (\ref{condition1}) and (\ref{condition2}), we find the following equation,
 \begin{equation}\label{equation1}
c_{1}V_{c}^{10}+c_{2}V_{c}^{8}+c_{3}V_{c}^{6}+c_{4}V_{c}^{4}++c_{5}V_{c}^{2}+b=0,
 \end{equation}
where we defined
\begin{eqnarray}
&&c_{1}=(9\times6^{(2/3)}\pi).\ \ c_{2}=(-3\times108 \pi^{(5/3)} Q^{2} + 36 \pi^{(2/3)})\nonumber\\
&&c_{3}=(48(6\pi)^{(1/3)}+ 432\times6^{(1/3)}\pi^{(7/3)} Q^{4}\nonumber\\
&&-3\times72\times6^{(1/3)}\pi^{(4/3)}Q^2 ).\nonumber\\
&&c_{4}=(32\times6^{(2/3)}+ 576\times6^{(2/3)}\pi^{2} Q^4\nonumber\\
&&+ 9\times6^{(2/3)}\pi\times(-32 Q^2)).\nonumber\\
&&c_{5}=(36\times\pi^{(2/3)}\times(-32 Q^{2}) + 108\times\pi^{(5/3)} Q^2\times(32 Q^2))\nonumber\\
&&b=(72\times6^{(1/3)}\times\pi^{(4/3)} Q^2 (32 Q^2)).
\end{eqnarray}
So, the critical volume is dependent on $ Q $.
\subsection{Stability}
\begin{figure}
\resizebox{0.40\textwidth}{!}
{%
\includegraphics{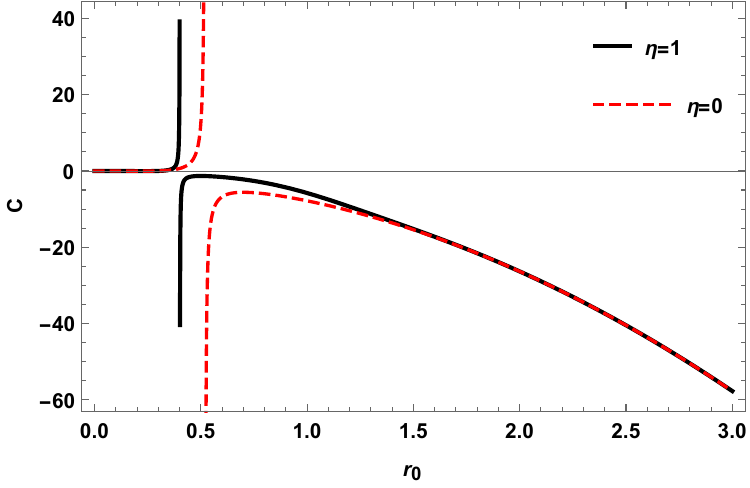}}
\vspace{0.5cm}       
\caption{Specific heat of RN black hole versus $r_{0}$ with $ Q=0.3 $ for $\eta=1$ and $ \eta=0 $.}
\label{heat2RN}
\end{figure}

In the presence of the non-perturbative correction, the specific heat of the RN black hole is obtained as,
\begin{equation}\label{specefic2}
C=\frac{2\pi r^{2}(r^{2}-Q^{2})(1-\eta e^{-S_{0}})^{2}}{(3Q^{2}-r^{2})+(r^{2}-2\pi r^{4}+Q^{2}(-3+2\pi r^{2}))\eta e^{-S_{0}}}
\end{equation}
In Fig. \ref{heat2RN}, we plotted it for $ \eta=1 $ and $ \eta=0 $.   { As the black hole size diminishes due to Hawking radiation \cite{I1}, an unstable/stable phase transition becomes evident. In the case of the $4D$ Reissner-Nordstr"{o}m black hole, a second-order phase transition occurs. This quantum correction induces a shift in the phase transition point of the black hole, highlighting the effect of quantum corrections on such transitions.}

\section{$4D$ Schwarzschild-AdS Black Hole}\label{SABH}
Schwarzschild-AdS (SAdS) black hole in four-dimensional space-time is given by the metric (\ref{metric-RN}) with the following metric function \cite{mann},
\begin{equation}\label{metric3}
f(r)=1-\frac{2M}{r}+\frac{r^{2}}{l^{2}},
\end{equation}
where the constant $l$ is the AdS curvature. The mass of the black hole, Hawking temperature and the black hole entropy are:
\begin{equation}\label{temperature3}
M=\frac{r_{0}}{2}(1+\frac{r_{0}^{2}}{l^{2}}),\ \ \ T_{H}=\frac{l^{2}r_{0}+3r_{0}^{3}}{4\pi r_{0}^{2}l^{2}},\ \ \ S_{0}=\pi r_{0}^{2}.
\end{equation}
Now, if we choose
\begin{equation}\label{metric modify3}
f(r)=1-\frac{2M^{\prime}}{r}+\frac{r^{2}}{l^{2}},
\end{equation}
and put $ M^{\prime}\longrightarrow(1-\alpha)M $ where $ \alpha=\eta e^{-S_{0}} $ then with (\ref{temperature-th-general-special}) we have the corrected temperature
\begin{equation}\label{Temperature modify3}
 T_{H}=\frac{l^{2}r_{0}+3r_{0}^{3}}{(1-\alpha)4\pi r_{0}^{2}l^{2}},
\end{equation}
which yields to the exponential corrected entropy (\ref{entropy modify2}).
\subsection{Thermodynamic quantities}
Helmholtz free energy is calculated as follows,
\begin{equation}\label{Helm3}
F=
\frac{r_{0}^{2}-l^{2}}{4l^{2}\pi r_{0}}+\frac{\eta}{8l^{2}\pi r_{0}}
\left(e^{-\pi r_{0}^{2}}2l^{2}+3r_{0}Erf[\sqrt{\pi}r_{0}]\right),
\end{equation}
while pressure is
\begin{eqnarray}\label{EOSr3}
P&=&\left( \frac{\partial F}{\partial V}\right)_{T}\nonumber\\
&=&\frac{\pi r_{0}^{2}(3r_{0}^{2}-l^{2})}{4l^{2}\pi r_{0}^{2}}
+\frac{\eta e^{-\pi r_{0}^{2}}(3r_{0}^{2}-l^{2})}{4l^{2}\pi r_{0}^{2}}.
\end{eqnarray}
Also, Gibbs free energy is obtained as,
\begin{eqnarray}\label{Gibbs3}
&&G=
\frac{r_{0}^{2}-l^{2}}{4l^{2}\pi r_{0}}+\frac{\pi r_{0}^{2}(3r_{0}^{4}+l^{2}(-r_{0}^{2}))}{3l^{2}r_{0}}\nonumber\\
&&+\frac{\eta}{8l^{2}\pi r_{0}}
\left(e^{-\pi r_{0}^{2}}2l^{2}+(3+2l^{2}\pi (1))r_{0}Erf[\sqrt{\pi}r_{0}]\right)\nonumber\\
&&+\frac{\eta e^{-\pi r_{0}^{2}}(3r_{0}^{4}-l^{2}r_{0}^{2})}{3l^{2}r_{0}}.
\end{eqnarray}
The specific heat for SAdS black hole is:
\begin{equation}\label{specefic3}
C=\frac{2\pi r_{0}^{2}(l^{2}+3r_{0}^{2})(1-\alpha)}{(-l^{2}+3r_{0}^{2})(1-\alpha)+2\pi r_{0}^{2}(l^{2}+3r_{0}^{2})(\alpha)}.
\end{equation}
We can discuss the stability of this kind in the next section.
\subsection{Critical points}
If we put $ r\longrightarrow\left( \frac{3V}{4\pi}\right) ^{\frac{1}{3}} $ and rescaling $ V^{\frac{1}{3}}\longrightarrow V $ in (\ref{EOSr3}) then we can find the following equation of state:
\begin{eqnarray}\label{EOS4}
P&=&\frac{a_{4}V^{2}(a_{2}V^{2}+a_{3}V^{4})}{288l^{2}\pi^{\frac{4}{3}}V^{6}}\nonumber\\
&+&4\eta\left( \frac{e^{-a_{5}V^{2}}(a_{2}V^{2}+a_{3}V^{4})}{288l^{2}\pi^{\frac{4}{3}}V^{6}}\right)
\end{eqnarray}
where
\begin{eqnarray}\label{EOS4-2}
a_{2}&=&-2(6\pi)^{\frac{2}{3}}l^{2},\ \ \ \ \ a_{3}=9\times 6^{\frac{1}{3}},\nonumber\\
a_{4}&=&6^{\frac{2}{3}}\pi^{\frac{1}{3}},\ \ \ \ \ \ \ \
a_{5}=\frac{1}{2}\times 3^{\frac{2}{3}}(\frac{\pi}{2})^{\frac{1}{3}}.
\end{eqnarray}
Applying the conditions (\ref{condition1}) and (\ref{condition2}), we find the following equation,
\begin{equation}\label{critical volume3}
c_{1}V_{c}^{10}+c_{2}V_{c}^{8}+c_{3}V_{c}^{6}+c_{4}V_{c}^{4}=0,
\end{equation}
where
\begin{eqnarray}
&&c_{1}=-27(6\pi)^{2/3},\nonumber\\
&&c_{2}=-\pi^{4/3}(28l^{2}),\nonumber\\
&&c_{3}=-6^{\frac{1}{3}}(28\pi l^{2}+90),\nonumber\\
&&c_{4}=-32(6\pi)^{2/3} l^{2}.
\end{eqnarray}
The critical volume is dependent on $l$.
\subsection{Joule-Kelvin expansion}
\begin{figure}
\resizebox{0.4\textwidth}{!}
{%
\includegraphics{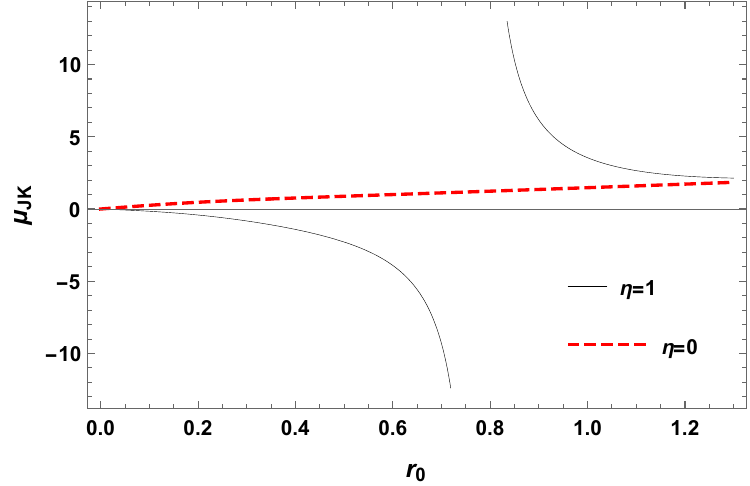}}
\vspace{0.5cm}       
\caption{Joule-Kelvin coefficient of SAdS black hole versus $r_{0}$ for $\eta=1$ and $ \eta=0 $ for $ l=0.5 $.}       
\label{heat4JKSAdS}
\end{figure}

\begin{figure}
\resizebox{0.4\textwidth}{!}
{%
\includegraphics{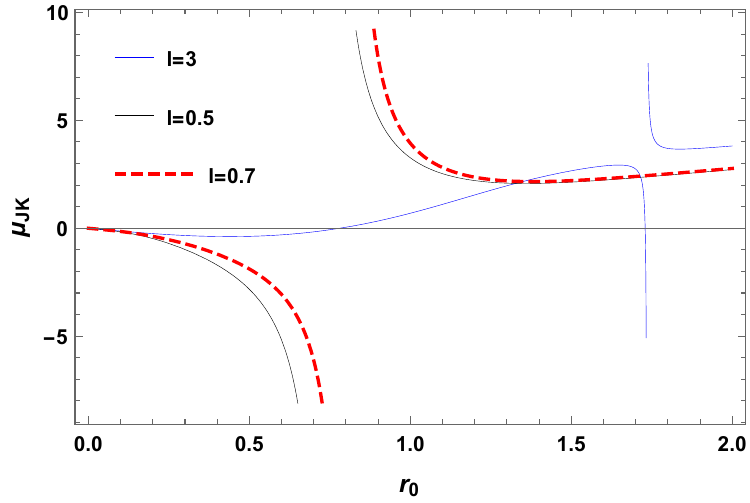}}
\vspace{0.5cm}       
\caption{Joule-Kelvin coefficient of SAdS black hole versus $r_{0}$ for $\eta=1$ and  different $l$.}     
\label{heat5JKSAdS}
\end{figure}

The Joule-Kelvin coefficient of SAdS black hole may be obtained via \cite{J1},
\begin{equation}\label{J-S7}
\mu_{JK}=\left(\frac{\partial T}{\partial P}\right)_{H}=\frac{1}{C_{p}}\left(T(\frac{\partial V}{\partial T})_{p}-V\right).
\end{equation}
  { Utilizing temperature and specific heat, the Joule-Kelvin coefficient can be determined from Fig. \ref{heat4JKSAdS} and Fig. \ref{heat5JKSAdS}. An intriguing observation is the emergence of asymptotic behavior in the Joule-Kelvin coefficient due to the effect of exponential correction. This behavior signifies how the coefficient behaves as the $4D$ Schwarzschild-AdS black hole temperature approaches zero or tends toward infinity. Notably, the value of the AdS curvature $l$ holds significant relevance in this context, as depicted in Fig. \ref{heat5JKSAdS}. This reveals the intricate relationship between the asymptotic behavior and the influence of $l$ on the coefficient dynamics.}

\section{$4D$ Charged AdS Black Hole}\label{CABH}
Charged AdS black hole in four-dimensional space-time is given by the metric (\ref{metric-RN}) with the following metric function \cite{mann},
\begin{equation}\label{metric4}
f(r)=1-\frac{2M}{r}+\frac{Q^{2}}{r^{2}}+\frac{r^{2}}{l^{2}}.
\end{equation}
where $l$ is the AdS radius.
The mass, Hawking temperature and entropy of a charged AdS black hole are given by
\begin{eqnarray}\label{temperature4}
M&=&\frac{r_{0}}{2}(1+\frac{Q^{2}}{r_{0}^{2}}+\frac{r_{0}^{2}}{l^{2}}),\nonumber\\
T_{H}&=&\frac{l^{2}(r_{0}^{2}-Q^{2})+3r_{0}^{4}}{4\pi r_{0}^{3}l^{2}},\nonumber\\
S_{0}&=&\pi r_{0}^{2}.
\end{eqnarray}
If we set
\begin{equation}\label{metric modify4}
f(r)=1-\frac{2M^{\prime}}{r}+\frac{Q^{2}}{r^{2}}+\frac{r^{2}}{l^{2}},
\end{equation}
with $ M^{\prime}\longrightarrow(1-\alpha)M $ where $ \alpha=\eta e^{-S_{0}} $ then with (\ref{temperature-th-general-special}) we have
\begin{equation}\label{Temperature modify4}
T_{H}=\frac{l^{2}(r_{0}^{2}-Q^{2})+3r_{0}^{4}}{(1-\alpha)4\pi r_{0}^{3}l^{2}}.
\end{equation}
So, the exponential corrected entropy is reproduced. It yields the following specific heat for a charged AdS black hole,
\begin{equation}\label{specefic4}
C=\frac{2\pi r_{0}^{2}(l^{2}(r_{0}^{2}-Q^{2})+3r_{0}^{4})(1-\alpha)}{X},
\end{equation}
where
\begin{eqnarray}
X&=&(l^{2}(-r_{0}^{2}+3Q^{2})+3r_{0}^{4})(1-\alpha)\nonumber\\
&+&2\pi r_{0}^{2}(l^{2}(r_{0}^{2}-Q^{2})+3r_{0}^{4})\alpha.
\end{eqnarray}

\begin{figure}
\resizebox{0.4\textwidth}{!}
{%
\includegraphics{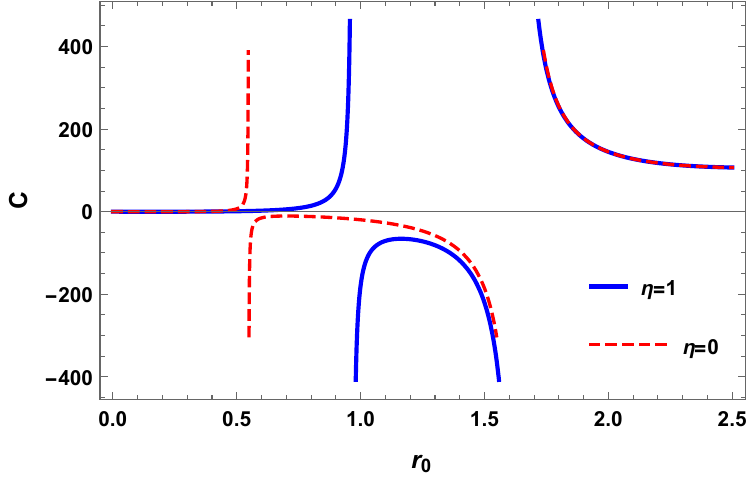}}
\vspace{0.5cm}       
\caption{Specific heat of the charged AdS black hole versus $r_{0}$ for $\eta=1$ and $ \eta=0 $ with $ Q=0.3 $ and $ l=3 $.}       
\label{heat4-CAdS}
\end{figure}
\begin{figure}
\resizebox{0.4\textwidth}{!}
{%
\includegraphics{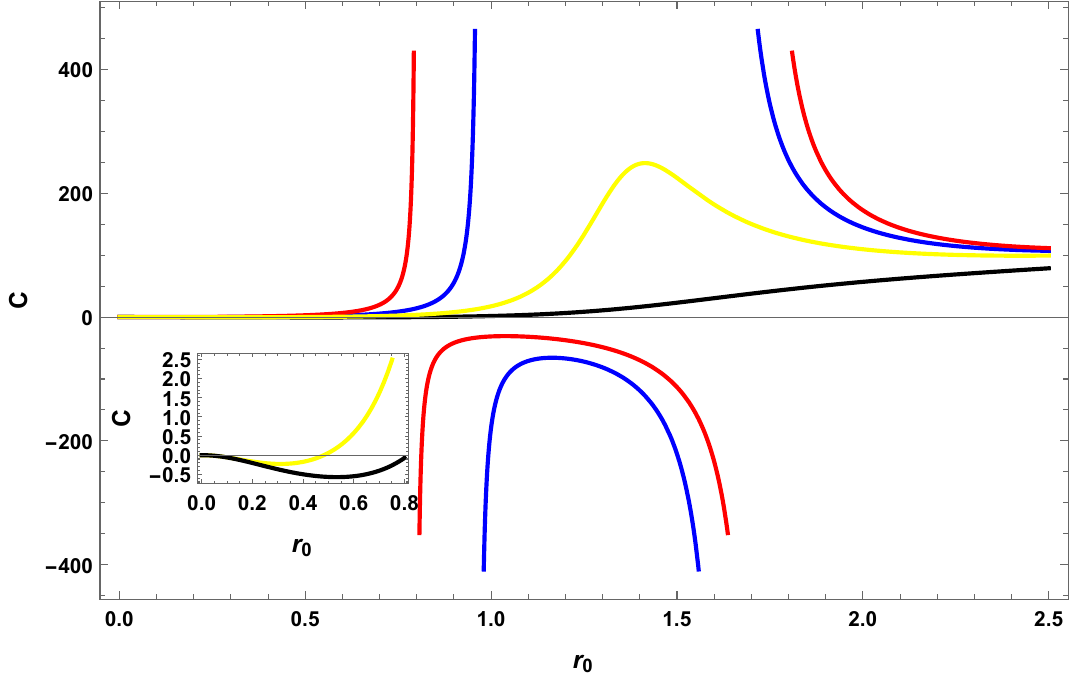}}
\vspace{0.5cm}       
\caption{Specific heat versus $r_{0}$ for $\eta=1$ and $ Q=0.09 $ (red), $ Q=0.3 $ (blue),  $ Q=0.5 $ (yellow) and $ Q=0.9 $ (black) for $ l=3 $.
For yellow curve, $ 0 <r_{0}<0.5 $ and black curve, $ 0 <r_{0}<0.8 $ the specific heat is negative.}
\label{differentheat4-CAdS}       
\end{figure}

 { In Figs. \ref{heat4-CAdS} and \ref{differentheat4-CAdS}, the behaviors of the specific heat for the charged AdS black hole are depicted. Fig. \ref{heat4-CAdS} demonstrates how the exponential correction alters the phase transition point. In Fig. \ref{differentheat4-CAdS}, the influence of the black hole charge on the specific heat is apparent. It reveals a second-order phase transition occurring at distinct points, highlighting the impact of quantum corrections on the $4D$ charged AdS black hole geometry. Notably, this divergent point might diminish for larger black hole charges, indicating the potential for complete stability in a strongly charged AdS black hole within four dimensions.}

\subsection{Thermodynamic quantities}
\begin{figure}
\resizebox{0.4\textwidth}{!}
{%
\includegraphics{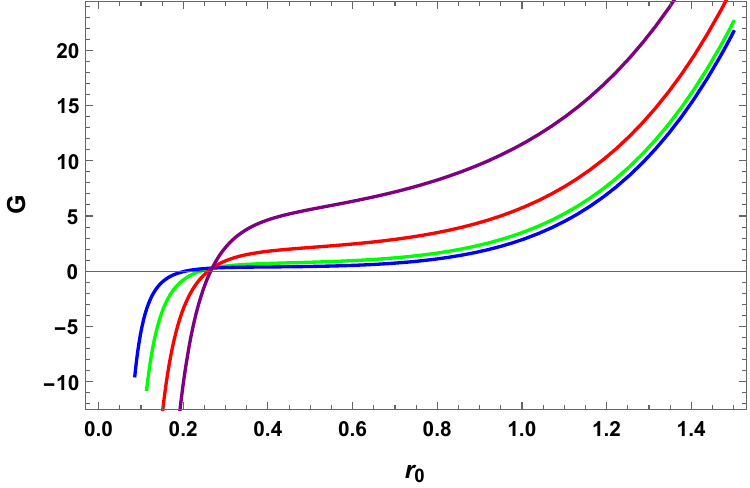}}
\vspace{0.5cm}       
\caption{Gibbs free energy the charged AdS black hole versus $r_{0}$ for $\eta=1$ and $ Q=0.3 $ (blue),  $ Q=0.5 $ (green),  $ Q=0.9 $ (red) and $ Q=1.5 $
purple with $ l=3 $.}
\label{Gibbs4-diagram-CAdS}       
\end{figure}

\begin{figure}
\resizebox{0.4\textwidth}{!}
{%
\includegraphics{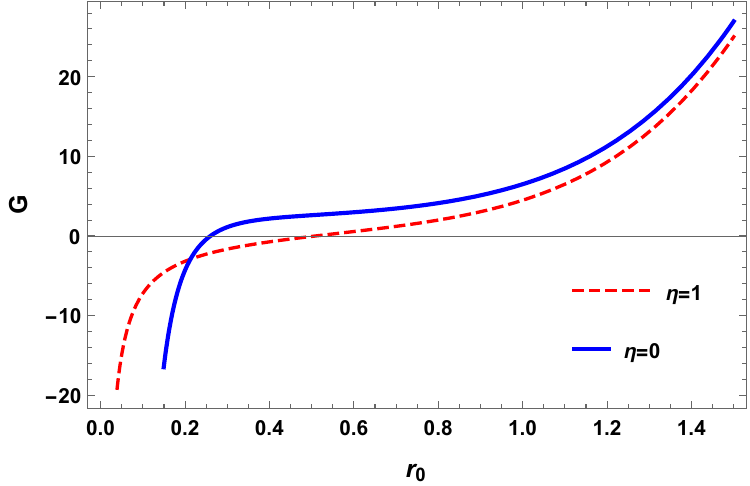}}
\vspace{0.5cm}       
\caption{Gibbs versus $r_{0}$ for $\eta=1$ and $\eta=0 $ for  $Q=0.3$  and $ l=3 $.}
\label{different Gibbs4-diagram-CAdS}
\end{figure}

Helmholtz free energy is calculated as follows,
\begin{eqnarray}\label{Helm4}
&&F=
\frac{2(r_{0}^{4}-l^{2}(3Q^{2}+r_{0}^{2}))}{8l^{2}\pi r_{0}^{3}}\nonumber\\
&&+\frac{\eta}{8l^{2}\pi r_{0}^{3}}
(e^{-\pi r_{0}^{2}}(r_{0}^{2}+Q^{2}(-1+2\pi r_{0}^{2}))2l^{2}\nonumber\\
&&+(3+2l^{2}\pi (1+2\pi Q^{2}))r_{0}^{3}Erf[\sqrt{\pi}r_{0}]),
\end{eqnarray}
which yields to the following pressure:
\begin{eqnarray}\label{EOSr4}
&&P=\left( \frac{\partial F}{\partial V}\right)_{T}=
\frac{\pi r_{0}^{2}(3r_{0}^{4}+l^{2}(3Q^{2}-r_{0}^{2}))}{4l^{2}\pi r_{0}^{4}}\nonumber\\
&&+\frac{\eta e^{-\pi r_{0}^{2}}(3r_{0}^{4}+l^{2}(3Q^{2}-r_{0}^{2}))}{4l^{2}\pi r_{0}^{4}}
\end{eqnarray}
Also, Gibbs free energy is calculated as
\begin{eqnarray}\label{Gibbs4}
&&G=
\frac{2(r_{0}^{4}-l^{2}(3Q^{2}+r_{0}^{2}))}{8l^{2}\pi r_{0}^{3}}+\frac{\pi r_{0}^{2}(3r_{0}^{4}+l^{2}(3Q^{2}-r_{0}^{2}))}{3l^{2}r_{0}}\nonumber\\
&&+\frac{\eta}{8l^{2}\pi r_{0}^{3}}
(e^{-\pi r_{0}^{2}}(r_{0}^{2}+Q^{2}(-1+2\pi r_{0}^{2}))2l^{2}\nonumber\\
&&+(3+2l^{2}\pi (1+2\pi Q^{2}))r_{0}^{3}Erf[\sqrt{\pi}r_{0}])\nonumber\\
&&+\frac{\eta e^{-\pi r_{0}^{2}}(3r_{0}^{4}+l^{2}(3Q^{2}-r_{0}^{2}))}{3l^{2}r_{0}}
\end{eqnarray}
In Fig. \ref{Gibbs4-diagram-CAdS}, we draw Gibbs free energy for several values of the black hole charge. We show that is an increasing function of the event horizon radius. Also, in Fig. \ref{different Gibbs4-diagram-CAdS}, we can see the effect of the exponential correction on the Gibbs free energy. As expected from a large radius, there is no important effect.

\subsection{Critical points}
\begin{figure}
\resizebox{0.4\textwidth}{!}
{%
\includegraphics{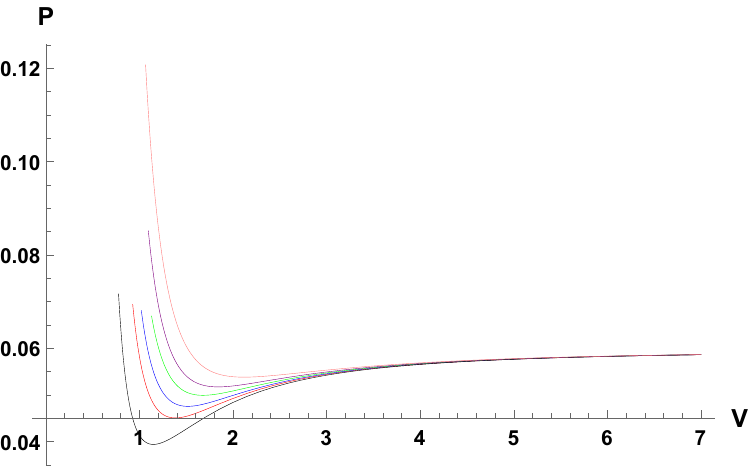}}
\vspace{0.5cm}       
\caption{Charged AdS black hole equation of state for $\eta=1$ and  $ Q=0.25 $ (Black), $ Q=0.31 $ (red), $ Q=0.35 $ (blue),  $ Q=0.4 $ (green), $ Q=0.45 $ purple and  $ Q=0.47 $ pink for $ l=3 $.}
\label{EOS5-CAdS}
\end{figure}

\begin{figure}
\resizebox{0.4\textwidth}{!}
{%
\includegraphics{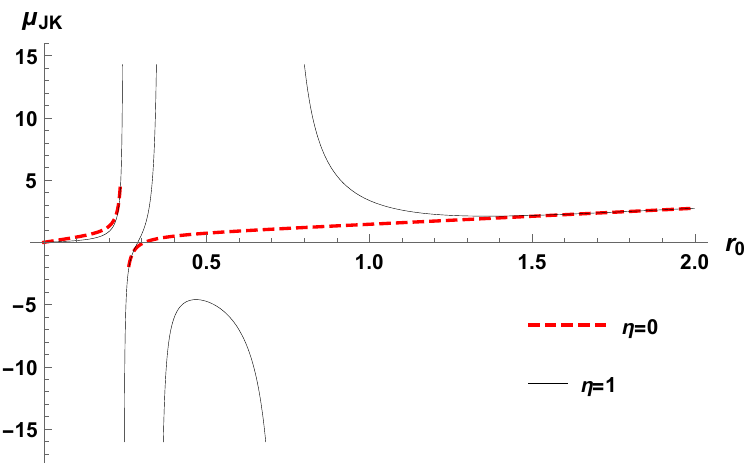}}
\vspace{0.5cm}       
\caption{Joule Kelvin coefficient versus $r_{0}$ for $\eta=1$ and $ \eta=0 $ for $Q=0.3$ and $l=0.9$.}
\label{18-SAdS1}       
\end{figure}

\begin{figure}
\resizebox{0.4\textwidth}{!}
{%
\includegraphics{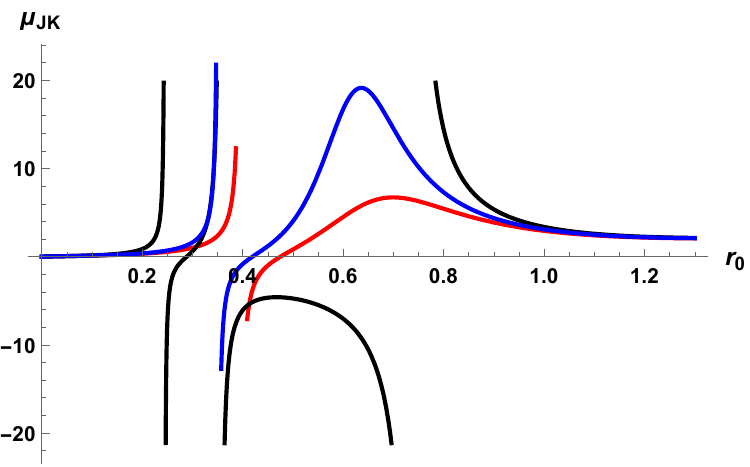}}
\vspace{0.5cm}       
\caption{Joule Kelvin coefficient versus $r_{0}$ for $\eta=1$ and $ Q=0.3 $ (black), $Q=0.5$ (Blue) and $Q=1$ (red) $l=0.9$.}
\label{18-SAdS2}      
\end{figure}

If we put $ r\longrightarrow\left( \frac{3V}{4\pi}\right) ^{\frac{1}{3}} $ and rescaling $ V^{\frac{1}{3}}\longrightarrow V $ in the equation (\ref{EOSr4}) then we can find the following equation of state:
\begin{eqnarray} 
P&=&\frac{a_{4}V^{2}(a_{1}Q^{2}+a_{2}V^{2}+a_{3}V^{4})}{288l^{2}\pi^{\frac{4}{3}}V^{6}}\nonumber\\
&+&4\eta\left( \frac{e^{-a_{5}V^{2}}(a_{1}Q^{2}+a_{2}V^{2}+a_{3}V^{4})}{288l^{2}\pi^{\frac{4}{3}}V^{6}}\right),
\end{eqnarray}
where
\begin{eqnarray}
a_{1}&=&24 \pi^{\frac{4}{3}}l^{2}\ \ \ \ \ a_{2}=-2(6\pi)^{\frac{2}{3}}l^{2}\ \ \ \ \ a_{3}=9\times 6^{\frac{1}{3}}\nonumber\\
a_{4}&=&6^{\frac{2}{3}}\pi^{\frac{1}{3}}\ \ \ \ \ \ \ \
a_{5}=\frac{1}{2}\times 3^{\frac{2}{3}}(\frac{\pi}{2})^{\frac{1}{3}}.
\end{eqnarray}
{In Fig. \ref{EOS5-CAdS}, we can see the behaviour of the corrected equation of state for various values of the black hole charge. It shows that a van der Waals equation of state may be a holographic dual picture of a charged AdS black hole in four dimensions.}
Applying the conditions (\ref{condition1}) and (\ref{condition2}), we find the following equation,
\begin{equation}\label{equation01}
c_{1}V_{c}^{10}+c_{2}V_{c}^{8}+c_{3}V_{c}^{6}+c_{4}V_{c}^{4}++c_{5}V_{c}^{2}+b=0,
\end{equation}
where we defined
\begin{eqnarray}
&&c_{1}=-27(6\pi)^{2/3},\nonumber\\
&&c_{2}=-\pi^{4/3}(28l^{2}+648Q^{2}),\nonumber\\
&&c_{3}=6^{\frac{1}{3}}(-28\pi l^{2}+422\pi Q^{2}+72\pi^{2}Q^{2}l^{2}\pi^{2}-90),\nonumber\\
&&c_{4}=(6\pi)^{2/3}(248\times Q^{2}+4Q^{2}l^{2}(-84+2\times 78\pi)\nonumber\\&&-32 l^{2}+72Q^{4}l^{2}\pi),\nonumber\\
&&c_{5}=78\times 24 l^{2}Q^{2}\pi^{4/3}+Q^{4}l^{2}\pi^{2}6^{\frac{1}{3}}\nonumber\\
&&\times4(2016-284\times 6^{2/3}\pi^{1/3} ),\nonumber\\
&&b=-2304\times6^{(1/3)}\pi^{2} Q^4 l^{2}.
\end{eqnarray}
Critical volume is dependent on $ Q $ and $l$.
Finally, the Joule-Kelvin coefficient \cite{J1},
\begin{equation}\label{J-S6}
\mu_{JK}=\left(\frac{\partial T}{\partial P}\right)_{H}=\frac{1}{C_{p}}\left(T(\frac{\partial V}{\partial T})_{p}-V\right),
\end{equation}
is drawn by Figs. \ref{18-SAdS1} and \ref{18-SAdS2}. { In the presence of the exponential correction, we can see one extra divergent point due to the modified geometry of black holes at quantum scales. There is also a local maximum in the Joule-Kelvin coefficient for the larger black hole charges, which may be due to the maximum changes in the black hole temperature.}

\section{STU Black hole}\label{STU}
The STU black hole may be given by the following line element \cite{STU1},
\begin{equation}\label{s1}
ds^{2}=-\frac{f(r)}{{\mathcal{H}}^{\frac{2}{3}}}dt^{2}
+{\mathcal{H}}^{\frac{1}{3}}(\frac{dr^{2}}{f(r)}+\frac{r^{2}}{R^{2}}d\Omega_{3}^{2}),
\end{equation}
where,
\begin{eqnarray}\label{s2}
f(r)&=&1-\frac{M}{r^{2}}+\frac{r^{2}}{R^{2}}{\mathcal{H}},\nonumber\\
{\mathcal{H}}&=&\prod_{i=1}^{3} H_{i},\nonumber\\
H_{i}&=&1+\frac{q_{i}}{r^{2}}, \hspace{10mm} i=1, 2, 3,
\end{eqnarray}
where $R$ is the constant AdS radius, while $r$ is the radial
coordinate along the black hole. Also, $q$ is the electric charge, and it is possible to consider $q_{1}=q_{2}=q_{3}=Q$ for simplicity. The thermodynamics of the STU black hole can be found in Ref. \cite{STU2}.\\
with comparing (\ref{s1}) with (\ref{metric-general}), we have:
\begin{equation}\label{coefficient}
g(r)=\frac{1}{\mathcal{H}^{\frac{2}{3}}},\ \ \ \ \ \ \ \ \frac{1}{h(r)}=\frac{\mathcal{H}^{\frac{1}{3}}}{f(r)}.
\end{equation}
Considering the STU black hole quantities, we have:
\begin{eqnarray}\label{MASS}
M&=&\frac{Q^{3}+3Q^{2}r_{0}^{2}+Q^{3}r_{0}^{4}+r_{0}^{6}+r_{0}^{4}R^{2}}{r_{0}^{2}R^{2}}\nonumber\\
T_{0H}&=&\frac{1}{2\pi}\sqrt{\frac{(Q^{3}(r_{0}^{4}-1)+r_{0}^{4}(2r_{0}^{2}+R^{2}))^{2}}{r_{0}^{4}(Q+r_{0}^{2})^{3}R^{4}}}\nonumber\\
S_{0}&=&\frac{1}{4GR^{3}}(r_{0}^{3}\sqrt{\mathcal{H}(r_{0})})
\end{eqnarray}
If we have:
\begin{equation} 
f(r)=1-\frac{2M^{\prime}}{r}+\frac{Q^{2}}{r^{2}}+\frac{r^{2}}{l^{2}}.
\end{equation}
and put $ M^{\prime}\longrightarrow(1-\alpha)M $ where $ \alpha=\eta e^{-S_{0}} $ then with (\ref{temperature-th-general}) we have the following corrected temperature,
\begin{equation}\label{Temperature modify5}
T_{H}=\frac{1}{2\pi(1-\alpha)}\sqrt{\frac{(Q^{3}(r_{0}^{4}-1)+r_{0}^{4}(2r_{0}^{2}+R^{2}))^{2}}{r_{0}^{4}(Q+r_{0}^{2})^{3}R^{4}}},
\end{equation}
by using the first law of thermodynamics (\ref{First law-general}), we can obtain the corrected entropy as,
\begin{equation}\label{entropy modify5}
 S=S_{0}+\eta e^{-S_{0}}
\end{equation}

\subsection{Thermodynamics}
\begin{figure}
\resizebox{0.4\textwidth}{!}
{%
\includegraphics{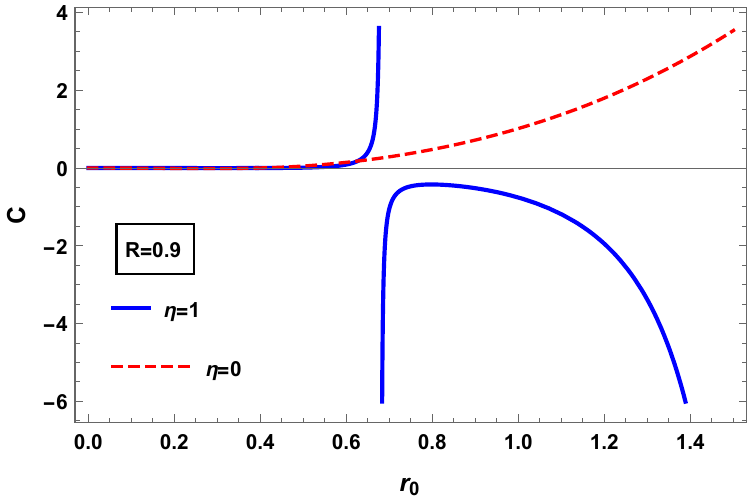}}
\vspace{0.5cm}       
\caption{Specific heat of STU black hole versus $r_{0}$ for $\eta=1$ and $ \eta=0 $ with $ Q=0.3 $ and $ R=0.9 $.}
\label{heat5stu1}       
\end{figure}

\begin{figure}
\resizebox{0.4\textwidth}{!}
{%
\includegraphics{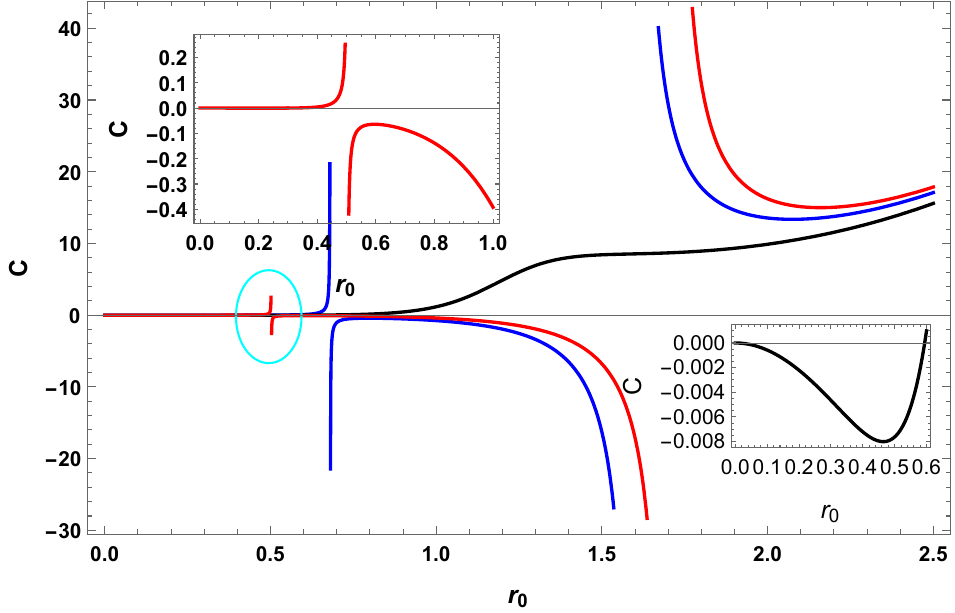}}
\vspace{0.5cm}       
\caption{Specific heat of STU black hole versus $r_{0}$ for $\eta=1$, $ Q=0.2 $ (red), $ Q=0.3 $ (blue) and $ Q=0.6 $ (black) and for $ R=0.9 $. }
\label{different heat5-stu}
\end{figure}

The specific heat for STU black hole is obtained as:
\begin{equation}\label{specefic4-stu}
C=\frac{\beta_{1}8\pi^{2}T_{0H}^{2}(1-\alpha)^{2}}{8\pi^{2}\beta_{1}\alpha T_{0H}^{2}+\beta_{2}(1-\alpha)},
\end{equation}
where
\begin{eqnarray}\label{specefic4-stu2}
\beta_{1}&=&\frac{S_{0}}{(Q+r_{0}^{2})}\nonumber\\
\beta_{2}&=&-4\pi^{2}T_{0H}^{2}(\frac{4}{r_{0}}+\frac{6r_{0}}{(Q+r_{0}^{2})})\nonumber\\
&+&\frac{4\pi T_{0H}}{r_{0}^{2}(Q+r_{0}^{2})^{5/2}R^{2}}.
\end{eqnarray}
In Fig. \ref{heat5stu1}, we can see the specific heat of the STU black hole to show that there is a phase transition point at small radii in the presence of the exponential correction. Also, in Fig. \ref{different heat5-stu}, we can see the effects of black hole charge and see for the black curve, $ 0 <r_{0}<0.6 $ specific heat is negative, which is a sign of instability.

\begin{figure}
\resizebox{0.4\textwidth}{!}
{%
\includegraphics{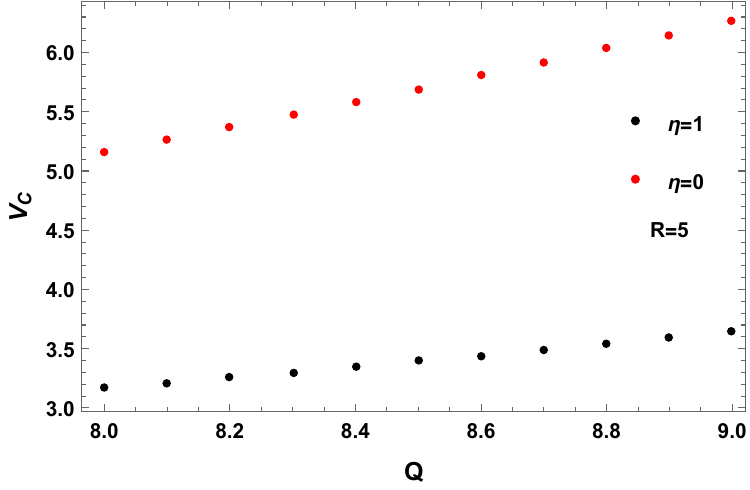}}
\vspace{0.5cm}       
\caption{Critical volume of STU black hole versus $Q$ for $\eta=1$ and $ \eta=0 $ for $ R=5 $.}
\label{13}       
\end{figure}

Helmholtz free energy is calculated as follows,
\begin{eqnarray}\label{helmholtz4}
&&F=\frac{(Q^{3}(-1+r_{0}^{4})+r_{0}^{4}(2r_{0}^{2}+R^{2}))}{T_{0H}R^{7}(32\pi^{2}r_{0}^{2})}\nonumber\\
&&(\frac{1}{r_{0}^{5}\sqrt{\frac{(Q+r_{0}^{2})^{3}}{r_{0}^{6}}}}(-Q^{3}(2+r_{0}^{4})+r_{0}^{4}(6Q+R^{2})\nonumber\\
&&+6Q^{2}r_{0}^{2}Log[r_{0}]+3Qr_{0}^{2}(-3Q+Q^{3}+R^{2}Log[Q+r_{0}^{2}])\nonumber\\
&&+\frac{\eta}{2R^{3}}(\frac{Q^{3}(143+28Q^{2})+10r_{0}^{6}+R^{2}(28Q^{2}-1120R^{4})}{35(Q+r_{0}^{2})}\nonumber\\
&&+\frac{560R^{6}(-3Q+Q^{3}+R^{2})}{35(Q+r_{0}^{2})^{2}}[\frac{1-Q^{2}-r_{0}^{2}}{Q+r_{0}^{2}}]\nonumber\\
&&+\frac{3r_{0}^{2}(16Q+21Q^{3}+21R^{2})}{35(Q+r_{0}^{2})}[-3r_{0}^{2}+Q])-\frac{(Q^{4}-16QR^{6})}{r_{0}^{2}(Q+r_{0}^{2})}\nonumber\\
&&+\frac{6Q^{7/2}Log[\frac{r_{0}}{Q+\sqrt{Q(Q+r_{0}^{2})}}]}{(Q+r_{0}^{2})^{3/2}}))
\end{eqnarray}\label{pressure4}
with the volume $V=\frac{r_{0}^{4}}{4R^{3}}(1+\frac{Q}{r_{0}^{2}})^{\frac{3}{2}}$ and the equation (\ref{helmholtz4}), we can find pressure:
\begin{eqnarray}
P&=&\frac{\omega_{1}\omega_{2}(Q+r_{0}^{2}+\sqrt{Q(Q+r_{0}^{2})})}{2(r_{0}^{3}(Q+r_{0}^{2})^{2}(\sqrt{Q}+\sqrt{Q+r_{0}^{2}})R^{2}(Q+4r_{0}^{2})}\nonumber\\
&&\times\left( 1-\frac{\eta}{4(Q+r_{0}^{2})^{3/2}R^{2}}\right)
\end{eqnarray}
where
\begin{eqnarray}
\omega_{1}&=&2Q^{4}+5Q^{3}r_{0}^{2}+2Q^{4}r_{0}^{4}+8Qr^{6}-Q^{3}r_{0}^{6}\nonumber\\
&+&2r_{0}^{8}+2Qr_{0}^{4}R^{2}-R^{2}r_{0}^{6},\nonumber\\
\omega_{2}&=&Q^{3}+3Q^{2}r_{0}^{2}+3Qr^{4}+r_{0}^{6}-16R^{6}.
\end{eqnarray}
We can find Gibbs free energy via the equation $ G=F+PV $.
Applying the conditions (\ref{condition1}) and (\ref{condition2}), we can find the numeric solution that depends on $ Q $ and $ R $. In Fig \ref{13}, we can see different critical volume for $\eta=0$ and $ \eta=1 $.

\subsection{Joule-Kelvin expansion}
\begin{figure}
\resizebox{0.4\textwidth}{!}
{%
\includegraphics{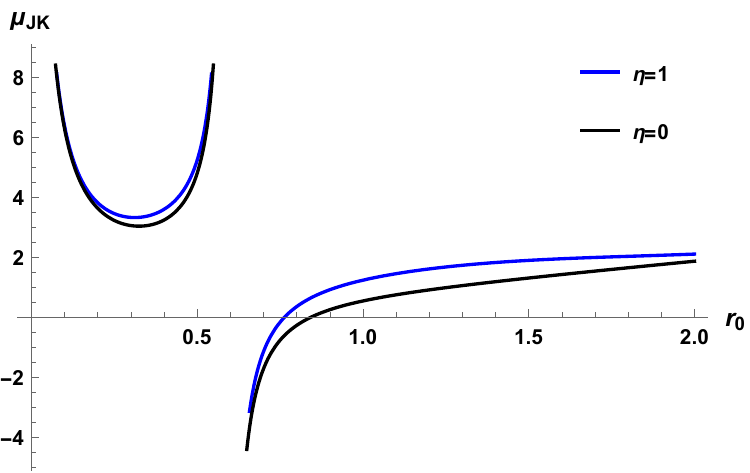}}
\vspace{0.5cm}       
\caption{Joule Kelvin coefficient of STU black hole versus $r_{0}$ for $\eta=1$ and $ \eta=0 $ for $ R=0.9 $ and $Q=0.6$.}
\label{16}       
\end{figure}

\begin{figure}
\resizebox{0.4\textwidth}{!}
{%
\includegraphics{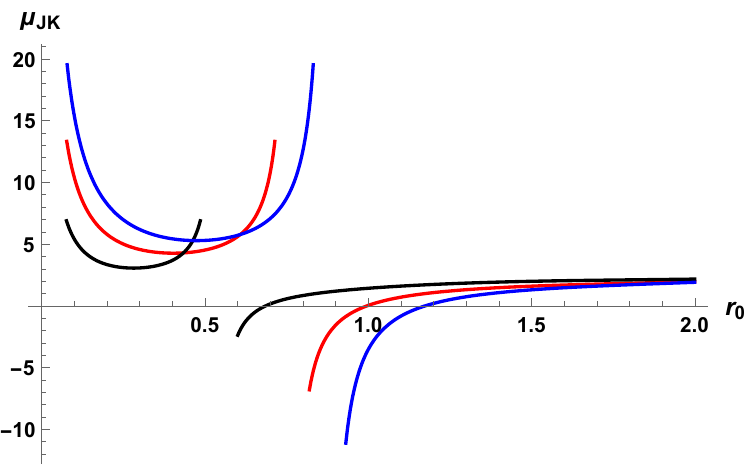}}
\vspace{0.5cm}       
\caption{Joule Kelvin coefficient of STU black hole versus $r_{0}$ for $\eta=1$ and $ Q=0.5 $ (black), $ Q=1 $ (red) and $Q=1.5$ (blue)  for $ R=0.9 $.}
\label{17}       
\end{figure}

The change of $E$ in an isothermal expansion is
\begin{equation} 
\triangle E=\int_{V_{1}}^{V_{2}}{\left[T\left(\frac{\partial P}{\partial T}\right)_{V}-P\right]dV}.
\end{equation}
Regarding cooling or warming, one can define the Joule-Kelvin coefficient given by the equation (\ref{J-S6}). So, the behaviour is drawn by Figs. \ref{16} and \ref{17}. We can see that at the small radii, the Joule-Kelvin coefficient is positive.

\newpage

\section{Conclusion}\label{C}
From a quantum perspective, black holes undergo mass reduction due to Hawking radiation \cite{I2,I3,I4,I5,I6}. This reduction may lead to either stable black holes or their complete evaporation. Understanding the quantum corrections impacting black hole thermodynamics is crucial in determining the final fate of a given black hole. Recent proposals introduced non-perturbative corrections to black hole entropy, notably an exponential term \cite{main}, which predominates at infinitesimal black hole sizes. The key inquiry revolves around the genesis of this exponential correction. 

In our study, we employed holographic principles and thermodynamic relations to derive modified metrics for select black hole solutions, facilitating the generation of corrected entropy that adheres to the proposed exponential correction \cite{main}. Our calculations demonstrate that by redefining the black hole mass, we can attain corrected black hole thermodynamics consistent with the findings seen in \cite{main}. Initially, our focus lay in deriving the corrected metric representative of a quantum black hole and subsequently evaluating its implications across various established black hole models. Our investigations underscored the profound influence of quantum effects on the structural aspects of black holes. Departing from conventional statistical methodologies, we leveraged holography to derive modified entropy, thereby enabling the acquisition of the corrected metric. This stands in contrast to existing literature, which predominantly concentrates on modifying entropy without delving into potential alterations in the black hole's spatial geometry. Our study firmly establishes that quantum effects indeed exert a tangible influence on the black hole's spatial configuration. Consequently, we affirm that the exponential correction significantly impacts the stability of black holes.

While the primary focus of this study was on non-spinning black holes, examining the impacts of rotating black holes presents an intriguing and unexplored area. Analyzing the modified metric in the context of rotating black holes could yield significant insights, especially in relation to the Banados, Silk, and West (BSW) mechanism \cite{B1,B2,B3}. Exploring how quantum corrections influence rotating black holes might unveil novel phenomena, providing deeper insights into the dynamics of these cosmic entities. This remains a prominent avenue for future research, poised to be on our immediate agenda.

{\section {Acknowledgements}
We sincerely thank the Editor and the anonymous Referees for their invaluable feedback and recommendations. \.{I}.S. would like to acknowledge networking support of COST Actions CA21106 and CA22113. He also thanks to T\"{U}B\.{I}TAK, ANKOS, and SCOAP3 for their support.
DVS thanks to the DST-SERB project (grant no. EEQ/2022/000824).}

\section*{Data Availability}
There is no associated data in this paper.

\end{document}